\documentclass[twocolumn,english,aps,prl,showpacs]{revtex4}
\usepackage[T1]{fontenc}
\usepackage[latin9]{inputenc}
\usepackage{amstext}
\usepackage{amssymb}
\usepackage{amsmath}
\usepackage{graphicx}
\usepackage{esint}

\makeatletter
\@ifundefined{textcolor}{}
{%
 \definecolor{BLACK}{gray}{0}
 \definecolor{WHITE}{gray}{1}
 \definecolor{RED}{rgb}{1,0,0}
 \definecolor{GREEN}{rgb}{0,1,0}
 \definecolor{BLUE}{rgb}{0,0,1}
 \definecolor{CYAN}{cmyk}{1,0,0,0}
 \definecolor{MAGENTA}{cmyk}{0,1,0,0}
 \definecolor{YELLOW}{cmyk}{0,0,1,0}
 }
\makeatother
\usepackage{babel}

\begin{document}

\title{Robust Photon Entanglement via Quantum Interference in Optomechanical Interfaces}

\author{Lin Tian}

\email{LTian@ucmerced.edu}

\affiliation{University of California, Merced, 5200 North Lake Road, Merced, California 95343, USA}


\begin{abstract}
Entanglement is a key element in quantum information processing. Here, we present schemes to generate robust photon entanglement via optomechanical quantum interfaces in the strong coupling regime. The schemes explore the excitation of the Bogoliubov dark mode and the destructive quantum interference between the bright modes of the interface, similar to electromagnetically induced transparency, to eliminate leading-order effects of the mechanical noise. Both continuous-variable and discrete-state entanglements that are robust against the mechanical noise can be achieved. The schemes can be used to generate entanglement in hybrid quantum systems between e.g. microwave photon and optical photon.
\end{abstract}
\maketitle

The mechanical modes in optomechanical systems \cite{reviews} can couple with cavity photons of wide range of frequencies as was demonstrated in recent experiments \cite{strongcouplingExp1, strongcouplingExp2, Groundstate5, EIT1, EIT2, EIT3, EIT4, Groundstate1, Groundstate2, Groundstate3, Groundstate4, Groundstate6, recentexp1, recentexp2}. Such systems can hence serve as an interface in hybrid quantum networks to connect optical and microwave photons \cite{quantumnetwork}. Quantum state transfer via such interfaces has been intensively studied \cite{pulsetransfer1, pulsetransfer2, stateconversion1, stateconversion2, stateconversion3, stateconversion4, stateconversion5, stateconversion6, stateconversion7, stateconversion8, stateconversion9, recentexp3}. The optomechanical interfaces have also been studied for entanglement generation between e.g. two cavity modes,  one cavity mode and one mechanical mode, or two mechanical modes \cite{ent2mode01, ent2mode02, ent2mode1, ent2mode2, ent2mode3, ent3mode01, ent3mode02, ent3mode1, ent3mode2, ent3mode3, ent3mode4, entother1, entother2, entother3, entother4, entother5, entother6, entother7, entpulse1, entpulse2, entpulse3, entpulse4}. The entanglement generated in those schemes is often limited by various factors such as the stability conditions that place constrains on the magnitude of the effective optomechanical couplings \cite{ent2mode1, ent2mode2, ent2mode3} and the amplification effect in the unstable regime \cite{entpulse2, entpulse3, entpulse4}. In particular, the thermal noise of the mechanical modes can strongly impair the entanglement.

The strong coupling regime where the effective optomechanical coupling exceeds the cavity bandwidth has recently been demonstrated in both microwave and optical cavities \cite{strongcouplingExp1, strongcouplingExp2, Groundstate5}. It hence becomes a practical objective to generate strong continuous-variable entanglement that can realize quantum teleportation with a fidelity exceeding the no-cloning boundary \cite{ent3mode4, cvent}. Here, stimulated by the experimental results, we present schemes to generate strong entanglement between photon modes that is robust against the mechanical noise. 

For photon modes that interact via a parametric Hamiltonian $H_{s}=-g_{s} (a_{1}a_{2}+a_{1}^{\dagger}a_{2}^{\dagger})$ with the coupling $g_{s}$, where $a_{i}$ ($i=1,2$) is the annihilation operator for mode $i$, continuous-variable entanglement can be generated \cite{cvent}. When applied to the vacuum state $\left|0_{1}0_{2}\right\rangle $, this Hamiltonian generates a two-mode squeezed vacuum state with entanglement $E_{N}=2r\log_{2}(e)$ quantified by the logarithmic negativity \cite{entln}. Under this interaction, the cavity operators at time $t$ can be written as $a_{i}(t)=\beta_{i}$, where
\begin{subequations}
\begin{align}
\beta_{1}&= \cosh(r)a_{1}+i\sinh(r)a_{2}^{\dagger}\label{eq:psib1} \\
\beta_{2}&= \cosh(r)a_{2}+i\sinh(r)a_{1}^{\dagger}\label{eq:psib2}
\end{align}
\end{subequations}
are the so-called Bogoliubov modes with a squeezing parameter $r=g_{s}t$. In our system, the cavity modes only couple with the mechanical mode and do not couple with each other. Entanglement between the photons is generated via their coupling with the mechanical mode which can induce strong mixing between the cavity and the mechanical components and expose the entanglement to the mechanical noise. In this work, exploring the excitation of the Bogoliubov dark mode and the quantum interference between the bright modes in an optomechanical interface, similar to electromagnetically induced transparency (EIT), we show that the cavity modes at selected time or cavity outputs at selected frequency can recover the Bogoliubov-like form defined in Eqs.(\ref{eq:psib1},\ref{eq:psib2}) with the leading-order mechanical components eliminated. The entanglement generated in these schemes is hence robust against the mechanical noise. In addition, we show that robust entanglement can also be achieved in discrete photon states via this interface. Compared with several previous works \cite{ent3mode01, ent3mode02, ent3mode1, ent3mode2, ent3mode3, ent3mode4}, this work studied the effect of the mechanical noise systematically and presented the conditions for robust entanglement generation in both cavity states and cavity outputs in the strong coupling regime. Our results show that the optomechanical interfaces can act as a noise-resilient hub in hybrid quantum networks to perform quantum state transfer and entanglement generation, which can facilitate the implementation of scalable hybrid systems. The schemes can be extended to similar systems such as two cavity modes coupling with a noisy qubit to implement high-fidelity quantum operations. 

\begin{figure}
\includegraphics[clip,width=7.5cm,clip]{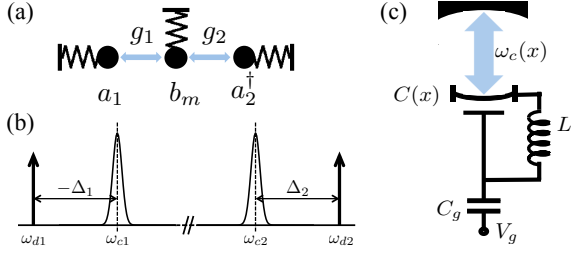}
\caption{(a) Three-mode model. (b) Spectrum of cavity resonances $\omega_{ci}$ and driving frequencies $\omega_{di}$ with $\Delta_{i}=\omega_{di}-\omega_{ci}$. (c) Schematic circuit of the interface. Frequency of the optical cavity $\omega(x)$ and capacitance of the microwave cavity $C(x)$ depend on mechanical displacement $x$. $L$: inductance of microwave cavity. $C_{g}$: gate capacitance. $V_{g}$: external drive.} 
\label{fig1}
\end{figure}

The optomechanical interface in our schemes is composed of two cavity modes coupling with a mechanical mode via the interaction $\sum\hbar G_{i}a_{i}^{\dag}a_{i}(b_{m}+b_{m}^{\dag})$ \cite{stateconversion4, stateconversion5, ent3mode3, ent3mode4}, where $b_{m}$ is the annihilation operator of the mechanical mode. One cavity is driven by red-detuned source with cavity detuning $\Delta_{1}$ to generate anti-Stokes processes and the other cavity is driven by blue-detuned source with cavity detuning $\Delta_{2}$ to generate Stokes processes, as is illustrated in Fig.\ref{fig1}(a,b). Let $\omega_{m}$ be the mechanical frequency and $-\Delta_{1}=\Delta_{2}=\omega_{m}$. With standard linearization, the effective Hamiltonian in the interaction picture of $H_{0}=\sum (-\hbar\Delta_{i}a_{i}^{\dagger}a_{i})+\hbar\omega_{m}b_{m}^{\dagger}b_{m}$ can be written as 
\begin{equation}
H_{I}=\hbar g_{1}(a_{1}^{\dagger}b_{m}+b_{m}^{\dagger}a_{1})+i\hbar g_{2}(a_{2}^{\dagger}b_{m}^{\dagger}-a_{2}b_{m})\label{eq:HI}
\end{equation}
where $g_{i}$'s ($i=1,2$) are the effective optomechanical couplings \cite{supple}. The environmental fluctuations can be represented by the cavity input operators $a_{in}^{(i)}(t)$ and the mechanical input operator $b_{in}(t)$. The correlation functions of the input operators are $\langle a_{in}^{(i)}(t)a_{in}^{(i)\dagger}(t^{\prime})\rangle=\delta(t-t^{\prime})$ and $\langle b_{in}(t)b_{in}^{\dagger}(t^{\prime})\rangle=(n_{th}+1)\delta(t-t^{\prime})$ in the high temperature limit with a thermal phonon number $n_{th}$ \cite{ent2mode2}. The Langevin equation for this system is
\begin{equation}
id\vec{v}(t)/dt=M\vec{v}(t)+i\sqrt{K}\vec{v}_{in}(t)\label{eq:Langevin}
\end{equation}
with $\vec{v}(t)=[a_{1}(t),b_{m}(t),a_{2}^{\dag}(t)]^{\textrm{T}}$ for the system operators, $\vec{v}_{in}(t)=[a_{in}^{(i)}(t),b_{in}(t),a_{in}^{(2)\dag}(t)]^{\textrm{T}}$ for the input operators, the diagonal matrix $K=\textrm{Diag}[\kappa_{1},\gamma_{m},\kappa_{2}]$, and 
\begin{equation}
M=\left [\begin{array}{ccc}
-i\frac{\kappa_{1}}{2} & g_{1} & 0\\
g_{1} & -i\frac{\gamma_{m}}{2} & ig_{2}\\
0 & ig_{2} & -i\frac{\kappa_{2}}{2}
\end{array}\right ],\label{eq:M}
\end{equation}
where $\kappa_{i}$'s and $\gamma_{m}$ are the cavity and the mechanical damping rates respectively. With $\omega_{m}\gg g_{i},\kappa_{i},\gamma_{i}n_{th}$, the rotating wave approximation has been applied above.
 
This model can be realized in many systems, e.g. the hybrid system of a microwave and an optical cavity coupling with a mechanical membrane as is shown in Fig.\ref{fig1}(c). We consider the system in the strong coupling regime with $g_{i}>\kappa_{i},\gamma_{m}$. The ratio $\kappa_{i}/g_{i}$ can reach $0.1$ for microwave cavities and $0.5$ for optical cavities \cite{strongcouplingExp2, Groundstate5}. Given the wide spectrum of possible systems, we will use arbitrary units for the model parameters but choose realistic ratios between these parameters in our discussion. The blue-detuned drive can induce instability in the interface which affects the entanglement generation. Using the Routh-Hurwitz criterion \cite{stability}, we derive the stability conditions for this model which can be approximated as $g_{1}^{2}/g_{2}^{2}>\max\left\{ \kappa_{2}/\kappa_{1},\kappa_{1}/\kappa_{2}\right\} $ in the strong coupling regime. This requires $g_{1}>g_{2}$ for the model to be stable. We can then write $g_{1}=g_{0}\cosh(r)$ and $g_{2}=g_{0}\sinh(r)$ with the squeezing parameter $r=\tanh^{-1}(g_{2}/g_{1})$. 

\emph{The Bogoliubov dark mode.} At zero damping, the eigenmodes $\alpha_{i}$ of the interface can be derived as
\begin{equation}
\alpha_{1}=\beta_{2}^{\dagger},\,\alpha_{2,3}=\left(\beta_{1}\pm b_{m}\right)/\sqrt{2}\label{eq:alpha0}
\end{equation}
with eigenvalues $\lambda_{1}=0$ and $\lambda_{2,3}=\pm g_{0}$ respectively. The mode $\alpha_{1}$ is related to the Bogoliubov mode defined in Eq.(\ref{eq:psib2}). This mode, which is called the Bogoliubov dark mode, is composed of the cavity modes only and is independent of the mechanical mode. Hence, the excitation of this mode is intrinsically exempted from the mechanical noise. The modes $\alpha_{2,3}$ are composed of both the cavity and the mechanical modes. These modes are hence subject to the mechanical noise and we call them the bright modes. An interesting feature of the bright modes is their symmetry. The superposition of the two bright modes yields the Bogoliubov mode defined in Eq.(\ref{eq:psib1}) with $(\alpha_{2}+\alpha_{3})/\sqrt{2}=\beta_{1}$, where the mechanical mode is eliminated. This superposition is then exempted from the mechanical noise as well. At finite damping, we treat the damping terms in the matrix $M$ as perturbations which modify the eigenmodes. We have
\begin{equation}
\alpha_{1}=\beta_{2}^{\dagger}+x_{1}b_{m},\,(\alpha_{2}+\alpha_{3})/\sqrt{2}=\beta_{1}-\sqrt{2}x_{3}b_{m}\label{eq:alphaN0}
\end{equation}
which contain first-order corrections $O(x_{j})b_{m}$ from the mechanical mode with $x_{j}=O(\kappa_{i}/g_{0},\gamma_{m}/g_{0})$ \cite{supple}. The eigenvalues are also modified by first-order imaginary parts as $\lambda_{1}=i\delta\lambda_{1}$ and $\lambda_{2,3}=\pm g_{0}+i\delta\lambda_{2}$, which strongly affect the entanglement in the cavity outputs.

The behavior of the optomechanical interface in both the time and the frequency domains is determined by the properties of these eigenmodes. Using the relations between the eigenmodes and the Bogoliubov modes, we will show below that cavity operators at selected time or cavity outputs at selected frequency can be exempted from the mechanical noise to the leading order.

\emph{Robust entanglement in cavity photons.} The time dependence of the cavity modes can be derived from the evolution of the eigenmodes. At zero damping, $\alpha_{1}(t)=\alpha_{1}(0)$ and $\alpha_{2,3}(t)=\exp(\mp i\varphi(t))\alpha_{2,3}(0)$ with the phase factor $\varphi(t)=\int_{0}^{t}dt^{\prime}g_{0}(t^{\prime})$. Applying Eq.(\ref{eq:alpha0}) to $\alpha_{i}(t)$, we derive $\beta_{2}(t)=\beta_{2}(0)$, which only contains the cavity modes, and
\begin{equation}
\beta_{1}(t)=\beta_{1}(0)\cos\varphi(t)-ib_{m}(0)\sin\varphi(t),\label{eq:beta1}
\end{equation}
which mixes the cavity and the mechanical modes. However,  at time $t_{n}$ with $\varphi(t_{n})=n\pi$ for integer $n$, we have $\beta_{1}(t_{n})=(-1)^{n}\beta_{1}(0)$. Both Bogoliubov operators at time $t_{n}$ hence only contain the cavity components and are exempted from the mechanical component $b_{m}(0)$ due to the destructive quantum interference between the mechanical components in $\alpha_{2,3}(t_{n})$ \cite{EITtheory1}. The cavity operators at time $t_{n}$ can then be derived from the Bogoliubov operators $\beta_{i}(t_{n})$ using Eqs.(\ref{eq:psib1},\ref{eq:psib2}).

For constant couplings at $t_{n}=n\pi/g_{0}$ for odd number $n$, we have $\beta_{1}(t_{n})=-\beta_{1}(0)$, and the cavity operators are 
\begin{equation}
\left [\begin{array}{c}
a_{1}(t_{n})\\
a_{2}^{\dag}(t_{n})
\end{array}\right ]=\left [\begin{array}{cc}
\cosh(2r) & -i\sinh(2r)\\
i\sinh(2r) & \cosh(2r)
\end{array}\right ]\left [\begin{array}{c}
-a_{1}(0)\\
a_{2}^{\dag}(0)
\end{array}\right ]\label{eq:atconstant}
\end{equation}
which generate a two-mode squeezed vacuum state with squeezing parameter $2r$ and entanglement $4r\log_{2}e$ when started from the vacuum state. Note for even number $n$, $a_{i}(t_{n})=a_{i}(0)$, and the cavities return to their initial states at $t_{n}$. For an adiabatic scheme with the couplings $g_{1}(t)= g_{0}\cosh(\lambda t)$ and $g_{2}(t)= g_{0}\sinh(\lambda t)$ under the condition $\lambda\ll g_{0}$ \cite{supple, LandauZener}, at $t_{n}=n\pi/g_{0}$ for integer $n$, 
\begin{equation}
\left [\begin{array}{c}
a_{1}(t_{n})\\
a_{2}^{\dag}(t_{n})
\end{array}\right ]=\left [\begin{array}{cc}
\cosh(r) & -i\sinh(r)\\
i\sinh(r) & \cosh(r)
\end{array}\right ]\left [\begin{array}{c}
(-1)^{n}a_{1}(0)\\
a_{2}^{\dag}(0)
\end{array}\right]\label{eq:atadiabatic}
\end{equation}
which generate a two-mode squeezed vacuum state with $r=\lambda t_{n}$. As the cavity operators at time $t_{n}$ do not contain the mechanical mode, the photon entanglement at $t_{n}$ is not subject to the influence of the thermal fluctuations in the initial mechanical state. 

At finite damping, the photon entanglement generated at time $t_{n}$ is affected by thermal fluctuations in both the initial mechanical state and the bath modes. Solving the Langevin equation \cite{supple}, we find that the cavity operators at $t_{n}$ include a term $O(x_{j})b_{m}(0)$ related to thermal fluctuations in the initial mechanical state and a term $O(\int dt^{\prime}\sqrt{\gamma_{m}}b_{in}(t^{\prime}))$ related to  thermal fluctuations in the bath modes. Let $n_{0}$ ($n_{th}$) be the thermal phonon number of the initial state (bath). These terms affect the covariance matrix of the cavity modes as $O(\kappa_{i}^{2}/g_{0}^{2})n_{0}$ and $O(\gamma_{m}/g_{0})n_{th}$ respectively. While at $t\ne t_{n}$, the cavity operators contain the mechanical mode as $O(1)b_{m}(0)$ which affects the covariance matrix as $O(1)n_{0}$. The destructive quantum interference between the eigenmodes at $t_{n}$ suppresses the thermal effects significantly by eliminating the leading order terms $O(1)b_{m}(0)$ from the cavity operators. The entanglement at $t_{n}$ is hence robust against the thermal noise. Note that the operators $a_{i}(t_{n})$ also include terms $O(x_{j})a_{i}(0)$ due to the decay of the eigenmodes and $O(\int dt^{\prime}\sqrt{\kappa_{i}}a_{in}^{(i)}(t^{\prime}))$ due to coupling to cavity bath, both of which affect the covariance matrix as $O(\kappa_{i}/g_{0})$. 
 
\begin{figure}
\includegraphics[clip,width=7.5cm]{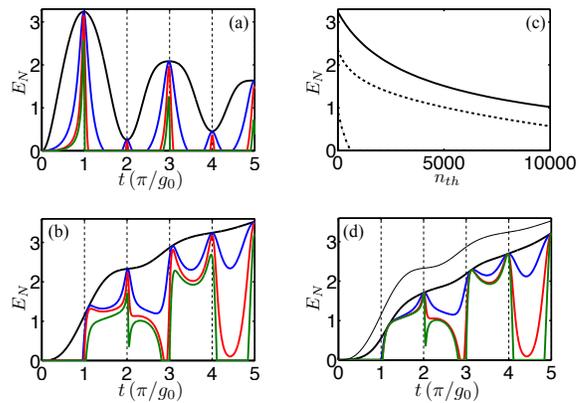}
\caption{(Color online) $E_{N}$ versus time for (a) constant couplings with $r=1$ and (b) adiabatic scheme with $r(t_{2})=1$. $n_{th}=0,10,10^{2},10^{3}$ from top to bottom. (c) $E_{N}$ versus $n_{th}$ for constant scheme at $t_{1}$ (solid), adiabatic scheme at $t_{2}$ (dashed), and stationary state scheme (dash-dotted). Above: $n_{0}=n_{th}$. (d) $E_{N}$ for adiabatic scheme. Thick: $n_{0}=0,10,10^{2},10^{3}$ and $n_{th}=10^{3}$ from top to bottom. Thin: $n_{0}=n_{th}=0$.}
\label{fig2}
\end{figure}
The entanglement is plotted in Fig.\ref{fig2}(a,b,c) for $n_{0}=n_{th}$ using numerical simulation \cite{supple}. The parameters we use are $g_{0}=3$, $(\kappa_{1},\kappa_{2}) =(0.3,02)$, and $\gamma_{m}=0.001$, all in arbitrary units and their ratios are within reach of current technology. Resonance peaks appear at $t_{n}$ and the peak values decrease slowly with $n_{th}$. As is shown Fig.\ref{fig2}(c), the entanglement at $t_{n}$ remains strong even for $n_{th}\sim10^{4}$, in sharp contrast to the stationary-state entanglement which quickly decreases to zero. In Fig.\ref{fig2}(d), we plot the entanglement for $n_{0}\ne n_{th}$ to distinguish the effects of the initial state noise and bath noise. It is shown that while the peak values decrease with the bath noise, the peak widths decrease with the initial state noise quickly. This is because the operators $a_{i}(t_{n}\pm\delta t)$ at  a small deviation $\delta t$ from $t_{n}$ contain the mechanical mode as $O(g_{0}\delta t)b_{m}(0)$, as can be derived from Eq.(\ref{eq:beta1}). This term affects the covariance matrix as $O(g_{0}^{2}\delta t^{2})n_{0}$ which can significantly narrow the peak widths for large $n_{0}$. The numerical results confirm our analytical results and show that robust photon entanglement can be generated at time $t_{n}$. 

\emph{Robust entanglement in cavity outputs.} Entangled photon pairs that are distributable in a quantum network are useful resource for quantum information processing \cite{cvent, ent3mode4}. Here we show that robust entanglement that survives high-temperature thermal noise can be generated in the cavity outputs of the optomechanical interface by appropriate frequency filtering. Define $a_{x}^{(i)}(\omega_{n})=\int d\omega g_{d}(\omega-\omega_{n})a_{x}^{(i)}(\omega)$ at $\omega_{n}=n\Delta\omega$ for integer $n$ and $x=in,out$, where $a_{x}^{(i)}(\omega)=\int dt a_{x}^{(i)}(t) e^{i\omega t}/\sqrt{2\pi}$ is the frequency component of the operator $a_{x}^{(i)}(t)$. For the simplicity of discussion, we use the filtering function $g_{d}(\omega) = 1/\sqrt{\Delta\omega}$ for $\omega\in(-\frac{\Delta\omega}{2},\frac{\Delta\omega}{2})$ and $g_{d}(\omega)=0$ otherwise. In experiments, more complicated filtering functions can be adopted \cite{ent3mode4}. The commutation relations between these operators are $[a_{x}^{(i)}(\omega_{m}),a_{x}^{(j)\dagger}(\omega_{n})]=\delta_{mn}\delta_{ij}$ which ensure that entanglement can be directly calculated from the covariance matrix of these operators \cite{cvent, supple}. 

\begin{figure}
\includegraphics[clip,width=7.5cm]{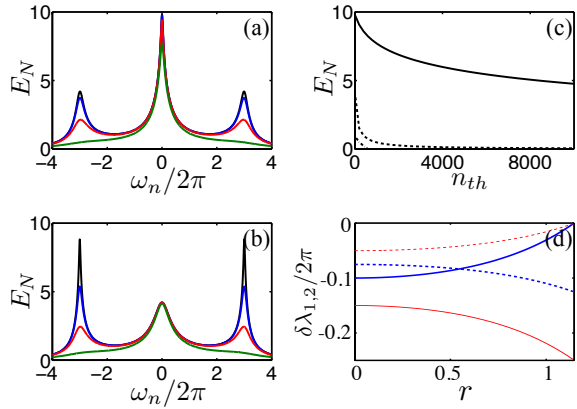}
\caption{(Color online) Entanglement versus $\omega_{n}$ for (a) $(\kappa_{1},\kappa_{2})=(0.3,0.2)$ and (b) $(0.2,0.3)$. $n_{th}= 0,10,10^{2},10^{3}$ from top to bottom. (c) $E_{N}$ versus $n_{th}$ at $\omega_{n}=0$ (solid), $\pm g_{0}$ (dashed), and stationary state scheme (dash-dotted) with $(\kappa_{1},\kappa_{2})=(0.3,0.2)$. (d) $\delta\lambda_{1}$ (solid) and $\delta\lambda_{2}$ (dashed) versus $r$. $(\kappa_{1},\kappa_{2})$: $(0.3,0.2)$ (thick blue); $(0.2,0.3)$ (thin red).}
\label{fig3}
\end{figure}

Using the Langevin equation and the input-output relation, we can derive the output operators in terms of the input operators. The entanglement between the cavity outputs can then be calculated, as is shown in Fig.\ref{fig3}(a,b) for the same $g_{0}$ and $\gamma_{m}$ as in Fig.\ref{fig2} and two sets of cavity damping rates. Three resonance peaks appear at $\omega_{n}=0,\,\pm g_{0}$, corresponding to strong excitation of three eigenmodes respectively. The peak widths are of the order of $|\delta\lambda_{i}|$ with $\delta\lambda_{i}\sim\kappa_{1,2}$ being the imaginary parts of the eigenvalues. As is illustrated in Fig.\ref{fig3}(c), the entanglement at $\omega_{n}=0$ decreases slowly with $n_{th}$ and is robust against the thermal noise; while the entanglement at $\omega_{n}=\pm g_{0}$ decreases quickly with $n_{th}$ to zero.

The excitations of the eigenmodes strongly depend on the frequency of the input modes. At $\omega_{n}=0$, the Bogoliubov dark mode is strongly excited with
\begin{equation}
\alpha_{1}=\left[\begin{array}{ccc}\sinh(r) & i x_{1} & i\cosh(r)\end{array}\right]\cdot\sqrt{K}\vec{v}_{in}(0)/\delta\lambda_{1}\label{eq:alpha1om0}
\end{equation}
and the bright modes are weakly excited with 
\begin{equation}
(\alpha_{2}+\alpha_{3})/\sqrt{2}=-\sqrt{\gamma_{m}}b_{in}(0)/g_{0}\label{eq:alpha23om0}
\end{equation}
and $\alpha_{2,3}\propto1/g_{0}$ in terms of $x_{1}$ and $\delta\lambda_{1}$ \cite{supple}. The excitations of the Bogoliubov modes, and hence the cavity modes and cavity outputs, can be derived using Eq.(\ref{eq:alphaN0}), which include the cavity inputs as $O(1/\sqrt{\kappa_{i}})a_{in}^{(i)}(0)$ and the mechanical input as $O(\sqrt{\gamma_{m}}/g_{0})b_{in}(0)$. The mechanical input is strongly suppressed due to the destructive quantum interference between $\alpha_{2}$ and $\alpha_{3}$. The ratio between the mechanical input and the cavity input contributions in the covariance matrix of the cavity outputs is $O(\kappa_{i}\gamma_{m}/g_{0}^{2})n_{th}$ where the dependence on $n_{th}$ is strongly suppressed. The entanglement in the cavity outputs is hence robust against the mechanical noise. At $\omega_{n}=g_{0}$ (and similarly at $\omega_{n}=-g_{0}$), the modes $\alpha_{1,3}$ are weakly excited with $\alpha_{1,3}\propto 1/g_{0}$. While the bright mode $\alpha_{2}$ is strongly excited with $\alpha_{2}\propto 1/\delta\lambda_{2}$ due to its resonance with the input frequency. The mechanical input terms in $\alpha_{2}$ and $\alpha_{3}$ cannot cancel due to their asymmetry at this frequency. The cavity outputs then include strong mechanical input terms which will strongly impair the entanglement as $n_{th}$ increases. 

The cavity outputs and entanglement strongly depend on $\delta\lambda_{1,2}$ which vary with the squeezing parameter $r$ and the damping rates $\kappa_{1,2}$. As is shown in Fig.\ref{fig3}(d), for $\kappa_{1}>\kappa_{2}$, $\delta\lambda_{1} \rightarrow0$ and $|\delta\lambda_{2}|$ becomes larger as $r$ increases towards the unstable regime, generating strong entanglement at $\omega_{n}=0$. For $\kappa_{2}>\kappa_{1}$, $\delta\lambda_{2} \rightarrow0$ and $|\delta\lambda_{1}|$ becomes larger as $r$ increases, generating strong entanglement at $\omega_{n}=\pm g_{0}$. Hence, to generate strong and robust entanglement for a hybrid interface with very different damping rates, we can choose the cavity mode with the larger damping rate to be mode $a_{1}$ by driving this mode with red-detuned source so that $\kappa_{1}>\kappa_{2}$. 

\emph{Robust entanglement in discrete states.} Quantum interference can also be used to generate robust discrete-state entanglement \cite{adiabatic}. Let the cavities both be driven by red-detuned sources with $-\Delta_{i}=\omega_{m}$. In \cite{stateconversion4,stateconversion5}, this setup was studied for high-fidelity quantum state transfer. Let $g_{1}(t)=g_{0}\sin(\lambda t)$ and $g_{2}(t)=-g_{0}\cos(\lambda t)$ vary adiabatically with $\lambda \ll g_{0}$. With $\lambda = g_{0}/4n$ for integer $n$, the cavity operators at the final time $t_{f}=\pi/4\lambda$ are
\begin{equation}
\left[\begin{array}{c}
a_{1}(t_{f})\\
a_{2}(t_{f})
\end{array}\right]=\frac{1}{\sqrt{2}}\left[\begin{array}{cc}
1 & -1\\
1 & 1
\end{array}\right]\left[\begin{array}{c}
a_{1}(0)\\
(-1)^{n}a_{2}(0)
\end{array}\right].\label{eq:atdiscrete}
\end{equation}
It can be proven that for the initial cavity state $|1_{1}0_{2}\rangle$, the final state of the cavities is $|\psi_{en}\rangle =(|1_{1}0_{2}\rangle +|0_{1}1_{2}\rangle)/\sqrt{2}$ \cite{supple}. Similarly, for the initial state $|0_{1}1_{2}\rangle$, the final state is $|\psi_{en}\rangle =(|1_{1}0_{2}\rangle -|0_{1}1_{2}\rangle)/\sqrt{2}$. The effect of the mechanical noise can also be studied by solving the Langevin equation. The cavity operators $a_{i}(t_{f})$ contain the mechanical mode as $O(\kappa_{i}/g_{0})b_{m}(0)$ which is suppressed by a factor $\kappa_{i}/g_{0}$ due to the destructive quantum interference between the eigenmodes. The discrete-state entanglement is thus robust against the mechanical noise. 

To conclude, we study an optomechanical interface for the generation of photon entanglement that is robust against the mechanical noise. Due to the excitation of the Bogoliubov dark mode and the quantum interference between the bright modes, the effect of the mechanical noise is significantly suppressed. Both continuous-variable and discrete-state entanglements can be achieved with realistic experimental parameters. When combined with the quantum state transfer schemes, this quantum interface provides a promising building block for hybrid quantum networks and for quantum state engineering. 

The author would like to thank Aashish Clerk, Ying-dan Wang, Sumei Huang, and Hailin Wang for very helpful discussions. This work is supported by DARPA ORCHID program through AFOSR, NSF-DMR-0956064, NSF-CCF-0916303, and NSF-COINS.

\end{document}